\documentclass[aps,pre,showkeys,showpacs,notitlepage,superscriptaddress]{revtex4-1}
\pdfoutput=1
\usepackage{amsmath}
\usepackage{amssymb}
\usepackage{amsfonts}
\usepackage{graphicx}
\usepackage{hyperref}
\usepackage{color}
\usepackage{multirow}
\usepackage{mathtools}
\usepackage{booktabs}
\usepackage{amsthm}

\hypersetup{pdftitle=Degree difference in complex networks}

\begin{document}
	\title{Degree difference: A simple measure to characterize 
		structural heterogeneity in complex networks}
	
	\author{Amirhossein Farzam}
	\email{farzam@mis.mpg.de}
	\affiliation{Max Planck Institute for Mathematics in the Sciences, Leipzig 04103 Germany}
	\author{Areejit Samal}
	\email{asamal@imsc.res.in}
	\affiliation{The Institute of Mathematical Sciences (IMSc), Homi Bhabha National Institute (HBNI), Chennai 600113 India}
	\affiliation{Max Planck Institute for Mathematics in the Sciences, Leipzig 04103 Germany}
	\author{J\"urgen Jost}
	\email{jost@mis.mpg.de}
	\affiliation{Max Planck Institute for Mathematics in the Sciences, Leipzig 04103 Germany}
	\affiliation{The Santa Fe Institute, Santa Fe, New Mexico 87501 USA}
	
	\begin{abstract}
		Despite the growing interest in characterizing the local 
		geometry leading to the global topology of networks, our 
		understanding of the local structure of complex networks, 
		especially real-world networks, is still incomplete. Here, we 
		analyze a simple, elegant yet 
		underexplored measure, `degree 
		difference' (DD) between vertices of an edge, to 
		understand the 
		local network geometry. We 
		describe the connection between 
		DD and global assortativity of the network from both formal and 
		conceptual perspective, and show that DD can reveal structural properties that are not 
		obtained from other such measures in 
		network science. Typically, edges with different DD play 
		different structural roles and the DD distribution is an 
		important network signature. Notably, DD is the
		basic unit of assortativity. 
		We provide an explanation as to why DD can 
		characterize structural heterogeneity in mixing patterns 
		unlike global assortativity and local node assortativity. By 
		analyzing synthetic and real networks, we show that DD 
		distribution can be used to distinguish between different 
		types of networks including those networks that cannot be 
		easily distinguished using degree sequence and global 
		assortativity. Moreover, we show DD to be an indicator for 
		topological robustness of scale-free networks. Overall, DD is 
		a local measure that is simple to define, easy to evaluate, and 
		that reveals structural properties of networks not readily seen 
		from other measures.
	\end{abstract}
	
	\maketitle
	
	\section*{Introduction}
	
	Since the dawn of network science \cite{Moreno1934,Moreno1943, 
		Newman1928}, scientists have tried to capture the structure and 
	dynamics of networks by measures that are simple to understand 
	and easy to evaluate (see e.g. \cite{Boccaletti2006,Newman2006,
		Caldarelli2007}). Early studies on the structure of complex 
	networks focused primarily on the global topology of these 
	discrete objects \cite{Watts1998,Barabasi1999,Albert2002}. 
	Global measures necessarily take some kind of average, and 
	therefore, such measures do not capture much of the individual 
	variability and heterogeneity in networks. To avoid this, it is 
	important to investigate local measures and their distributions
	in complex networks. 
	Local measures are in particular of substantial interest for 
	analyzing massive networks, where global measures are either 
	impractical to compute or fail to provide the desired information 
	about network components. It is thus natural to witness growing 
	interest among the network science community to investigate the 
	local geometry of complex networks (see e.g. 
	\cite{Bauer2012,Jost2014,Sandhu2015,Sreejith2016,Samal2018}).
	Local clustering coefficient \cite{Watts1998}, generalized degree, local 
	assortativity \cite{Piraveenan2008}, Ollivier-Ricci curvature 
	\cite{Jost2014,Sandhu2015,Samal2018,Ni2019} and Forman-Ricci 
	curvature \cite{Sreejith2016,Samal2018} are some of the notable 
	measures characterizing the local structural properties of 
	complex networks.

	Moreover, with the recent exception of discrete Ricci type 
	curvature measures, such local or global measures are typically 
	evaluated on vertices, rather than on edges, although the edges 
	are of course what really constitutes a network. In this work, 
	we shall therefore systematically pursue an edge-based approach 
	to characterize the local structure of complex networks. As 
	mentioned, discrete Ricci type curvature measures are local and 
	edge-based, and they are by now established as useful tools for 
	the analysis of empirical networks \cite{Sandhu2015,Sandhu2016,Sreejith2016,Samal2018,Weber2017}. For instance, the Forman-Ricci 
	curvature of an edge in an unweighted and undirected network 
	essentially evaluates the sum of the degrees of its two 
	vertices, and edges with a large such sum are important for the 
	cohesion of the network in question and therefore deserve 
	attention. However, when we want to understand the local 
	heterogeneity in a network, Forman-Ricci curvature may not be 
	so useful, because it does not distinguish between an edge that 
	connects two vertices of intermediate but similar degrees, from 
	an edge that connects a highly connected vertex with a sparsely 
	connected one; in both cases, the sum of their degrees is large. 
	Now, there is an important and well-established global concept 
	for judging the homogeneity or heterogeneity of a network, its 
	\textit{assortativity} (see for instance \cite{Jackson2008}). A 
	network is assortative if on average, the degrees of connected 
	vertices are similar, and disassortative, if they tend to be 
	rather different. For instance, many social networks, 
	particularly those formed through group-to-group connections, 
	are known to be assortative \cite{Fisher2017}, i.e. agents with 
	high degree seem to connect to other high-degree agents, and 
	similarly, low-degree agents tend to connect to agents 
	with lower rather than higher degree. Again, this property
	cannot be captured by node-based quantities, such as 
	the degree sequence, because a simple rewiring can transform 
	an assortative into a disassortative network or vice versa, 
	without changing the degree sequence. This motivates us to 
	systematically explore the `degree difference' between two 
	vertices of an edge in complex networks.
	
	Degree difference or closely related measures have been 
	previously used in the study of complex networks, for instance, 
	to study scale-free properties of networks \cite{Zhou2020},
	and to investigate the structure of inter-organizational networks 
	\cite{yang2010multilevel}.
	However, curiously, so far there seems to have been no systematic 
	analysis of degree difference in complex networks, although 
	for the reasons explained above, this is a most natural local 
	measure. It is simple to define, easy to evaluate, and captures 
	the local picture underlying assortativity or disassortativity.
	Moreover, we shall show in this contribution that the measure 
	provides novel insight into both synthetic and real networks. For 
	the synthetic networks, we will derive explicit formulae, thereby 
	laying the foundations for a theoretical investigation.
	
	In fact, assortativity can be defined more generally 
	\cite{Newman2010} to express how similar or dissimilar 
	neighbouring vertices are with respect to some quantity 
	$\alpha$. In particular for social networks \cite{Jackson2008}, 
	this is important as it connects to homophily, that is, the 
	tendency to associate with like-minded or otherwise similar 
	people. Thus, for a graph $G(V,E)$, with vertex set $V$ and 
	edge set $E$, and an attribute $\alpha: V \rightarrow{} 
	\mathbf{F}$, mapping the $n$ vertices in $V$ to elements in 
	$\mathbf{F}$, assortativity captures heterogeneity in mixing 
	patterns in $G$ at a global scale. The global assortativity 
	(GA) with respect to $\alpha$ is given by
	\begin{equation}
		r_{\alpha} = \cfrac{Tr(e) - \lVert{e^2}\rVert_{L_1}}{1 - 
			\lVert{e^2}\rVert_{L_1}}
		\label{eq:assortativity}
	\end{equation}   
	where $e$ is the $n \times n$ matrix of joint probabilities 
	with $e_{i,j} = P(\alpha(i), \alpha(j))$, $Tr(e)$ is the trace, 
	and $\lVert{e^2}\rVert_{L_1}$ is the $L_1$ norm of $e$. When 
	$\alpha$ maps the vertices to their degrees, we denote 
	$r_{\alpha}$ by $r$, and Eq. \ref{eq:assortativity} is equal to 
	the Pearson correlation coefficient of the degrees of connected 
	vertices. By convention, if the term \textit{assortativity} is 
	used without specifying the attribute, the attribute is assumed 
	to be the degree.
	
	There have been previous attempts to break assortativity down 
	to its more local components. Piraveenan \textit{et al} 
	\cite{Piraveenan2010} define a local point-wise measure of 
	assortativity, local node assortativity (LNA), denoted by 
	$\hat{\rho}_v$, quantifying the contribution of each vertex $v$ 
	to the GA in the network as follows
	\begin{equation}
		\hat{\rho}_v = \cfrac{j(j+1)(\bar{k}_v-\mu_q)}{2M{\sigma_q}^2}
		\label{eq:unbiased_local_assortativity}
	\end{equation}
	where $j$ is the excess degree of $v$, $\bar{k}_v$ is the
	average excess degrees of the neighbours of $v$, $\mu_q$ is the 
	global average excess degree, $M$ is the number of edges in the 
	network, and $\sigma_q$ is the standard deviation of the excess 
	degree distribution in the network. 
	Consider an edge $e=\{v, u\}$. As we arrive at $v$ via the edge 
	$e$, excess degree of $v$, also referred to as \textit{remaining degree} of $v$, is the number of neighbours of $v$ other than the 
	vertex $u$ from which we arrived at $v$. Given how excess degree 
	is defined, it is the natural measure to consider when comparing 
	the number of neighbours of two connected vertices, since the 
	connection between the two is a given and the connections other than 
	the edge between the two neighbours are those contributing to the 
	similarity or difference in their degrees. Hence, excess degree of 
	the vertices is the measure used for computing GA and LNA. Note that, 
	although conceptually excess degree of a vertex is defined with an 
	incident edge in mind, formally excess degree is, in fact, equal to 
	degree minus 1.
	
	GA can be obtained from LNA through the following identity
	\begin{align}
		r &= \sum_{j, k \in D(V)}\cfrac{jk(e_{j,k} - q_j q_k)} {\sigma_q ^2} 
		\nonumber \\
		&= \frac{1}{\sigma_q ^2} \bigg[ \big( \sum_{j,k \in D(V)} jk e_{j,k} 
		\big) - \mu_q ^2 \bigg] \label{eq:mid0_equn_lna_to_assort} \\
		&= \cfrac{1}{\sigma_q ^2} \ \sum_{j} \sum_{v \in V_j} \bigg( \Big[ 
		(j+1) \cfrac{j \bar{k}_v}{2M} \Big] - \Big[ (j+1) \cfrac{j\mu_q}
		{2M} \Big] \bigg) \label{eq:mid_equn_lna_to_assort} \\
		&= \cfrac{1}{\sigma_q ^2} \ \sum_{j} \sum_{v \in V_j} \hat{\rho}_v \nonumber
	\end{align}
	where $D(V)$ denotes the set of degrees of vertices in $V$, and 
	$V_j$ is the set of vertices in $V$ with excess degree $j$. In 
	Eq. \ref{eq:mid_equn_lna_to_assort}, the term within the first 
	braces is the contribution of vertex $v$ to the first term in 
	Eq. \ref{eq:mid0_equn_lna_to_assort} and the term within the 
	second braces is its contribution to $\mu_q ^2$.
	
\begin{figure}[htb]
		\includegraphics[width=.81\columnwidth]{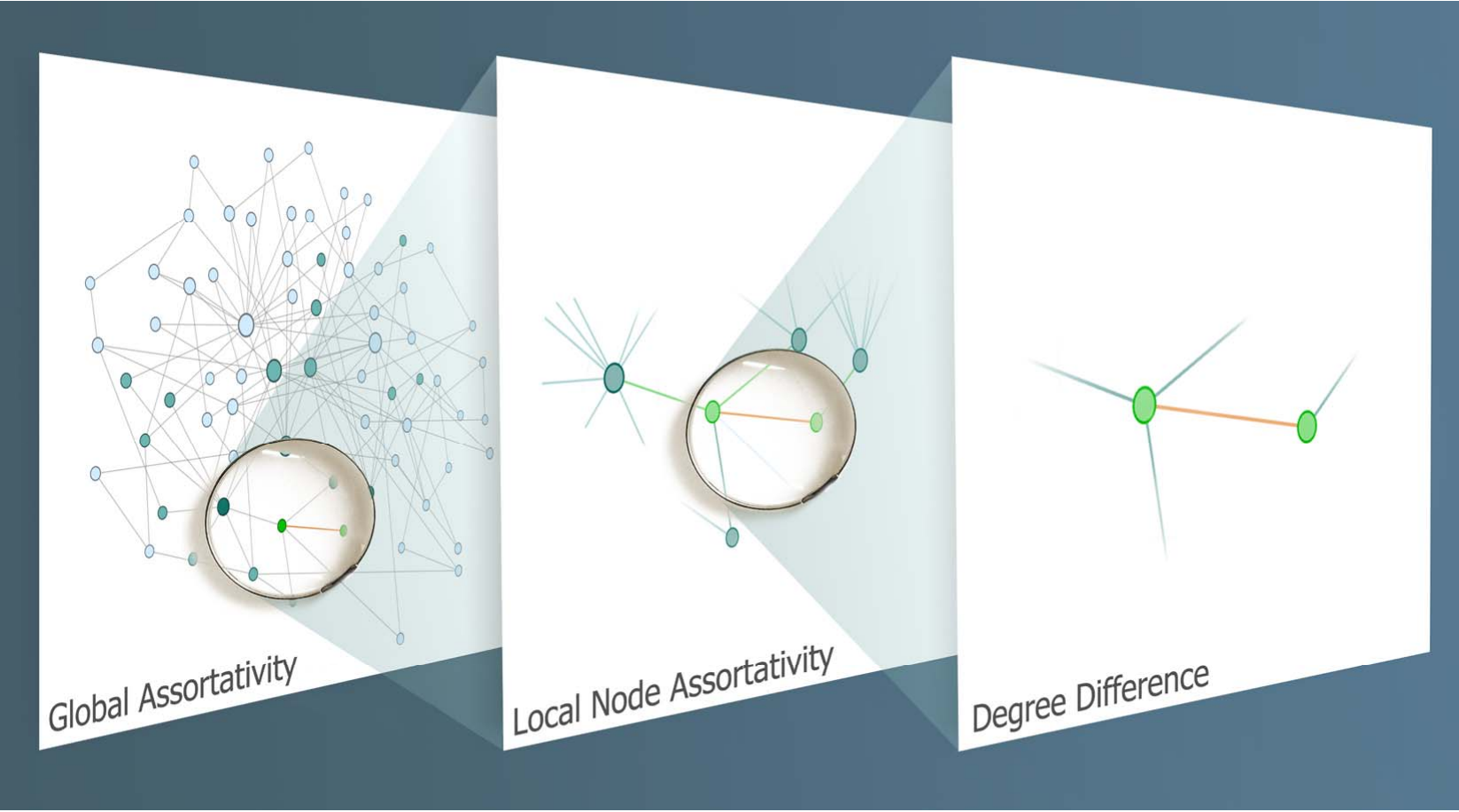}
		\caption{Heterogeneity in mixing patterns in networks is 
			quantified through global assortativity (GA) given by Eq. 
			\ref{eq:assortativity} at the global scale. At a more local scale, 
			mixing pattern is quantified through local node assortativity 
			(LNA) given by Eq. \ref{eq:unbiased_local_assortativity} which 
			aggregates the heterogeneity in degrees of the vertices anchoring 
			the edges incident on the vertex. Degree Difference (DD) is the 
			basic local unit contributing to assortativity 
			and captures heterogeneity in mixing 
			patterns at the scale of individual edges.}
		\label{fig1}
	\end{figure}

\begin{figure}[htb]
		\includegraphics[width=.81\columnwidth]{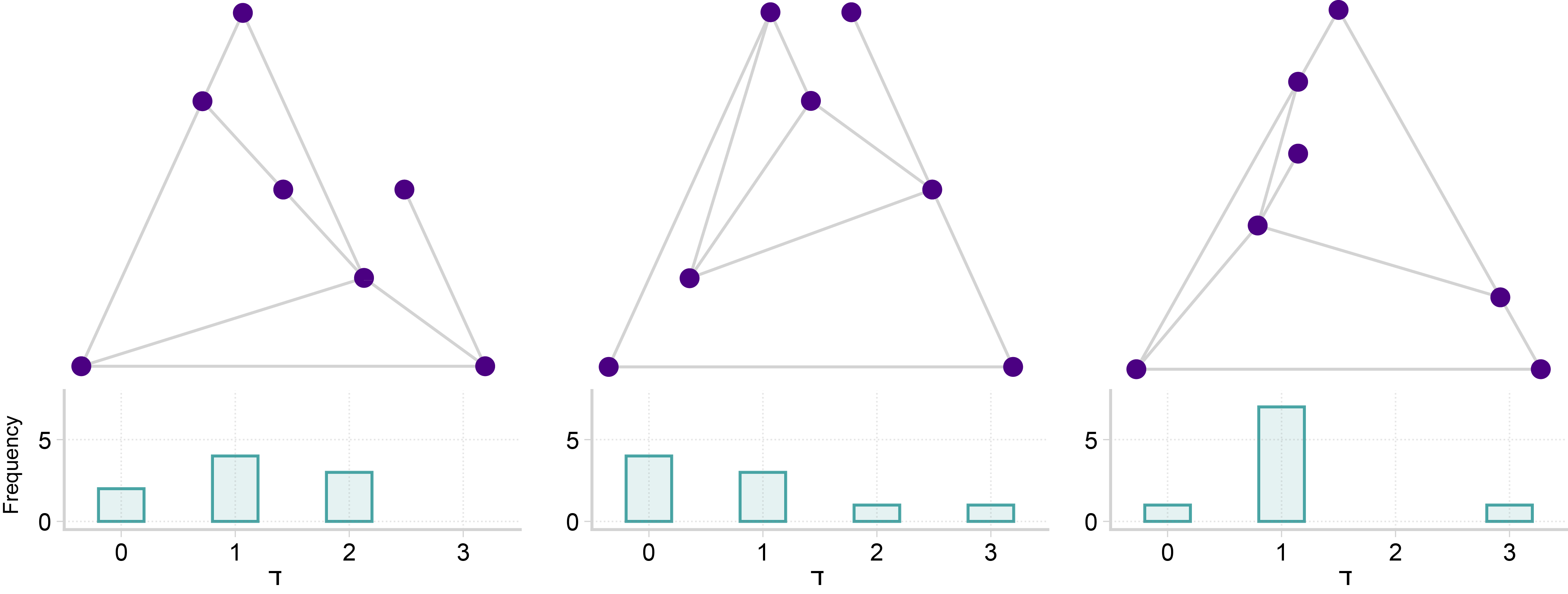}
		\caption{Graphs with same degree sequence and same global 
			assortativity (GA) can have different degree difference (DD)
			distributions. The figure shows 3 graphs with $7$ vertices and
			$9$ edges with same degree sequence and GA of $\approx -0.358$, 
			but with different DD distributions. In the figure, the bar plot 
			below each graph shows the DD distribution.}
		\label{fig2}
	\end{figure}

	While this represents a valuable step towards understanding 
	local mixing patterns in networks, LNA appears somewhat 
	complicated and is defined on the vertices. In fact, at first 
	sight it seems most natural to come up with such a node-based 
	measure. But recall that assortativity is evaluating 
	similarities or differences between neighbouring vertices, and 
	two neighbouring vertices are nothing but an edge. Therefore, 
	it seems more natural to evaluate quantities directly on edges. 
	That is, in fact, our starting point. Thus, we shall decompose 
	assortativity into its basic unit, the degree difference 
	(DD) between the vertices forming an edge. Given an edge $e=\{v,u\}$ 
	in an unweighted and undirected graph linking the vertices $v$ 
	and $u$ with degrees $\text{deg}(v)$ and $\text{deg}(u)$, 
	DD of $e$ is given by
	\begin{align}
		\daleth(e) &= \big| \text{deg}(v) - \text{deg}(u) \big|
		\label{eq:degdiff_unweighted}
	\end{align} 
	where $\daleth: E \rightarrow{} \mathbb{Z}^{\geq}$ is a 
	function from the edge set of the graph to non-negative 
	integers, mapping $e$ to absolute value of its DD.
	Similarly, for directed graphs, we define directed DD (diDD) as 
	follows
	\begin{align}
		\daleth_{\rightarrow{}}(e) &=  \text{deg}^{out}(u) - \text{deg}^{in}(v)
		\label{eq:di_degdiff_unweighted}
	\end{align} 
	where $e=(v, u)$ is the directed edge from $v$ to $u$ and
	$\daleth_{\rightarrow{}}: E \rightarrow{} \mathbb{Z}$ has the 
	entire set of integers as its codomain. 
	Note that there are four possible ways to define the directed DD 
	(diDD), corresponding to the four permutations of in- or out- 
	degree of the head vertex minus in- or out- degree of the tail 
	vertex. The variation given in Eq. \ref{eq:di_degdiff_unweighted} 
	is most consistent with the orientation on the edge and the 
	direction of potential flow.
	After verifying that this simple and elegant network measure 
	meaningfully captures structural similarities and differences, 
	we here show that DD is independently informative and capable 
	of characterizing network structure. 
	Importantly, DD of edges, through (indirect) quantification of 
	the contribution of individual edges to the GA, is the 
	basic unit of assortativity, as illustrated in 
	Figure \ref{fig1}.
	Furthermore, we provide an explanation as to 
	why DD can characterize structural heterogeneity in mixing 
	patterns, a feature that is lost due to averaging when 
	employing the measures GA or LNA. In Figure \ref{fig2}, we show 
	three graphs with same degree sequence and same GA that have 
	different DD distributions.

	The remainder of this paper is organized as follows. In the 
	next section, we derive the analytical formulae for DD 
	distribution in Erd\"{o}s-R\'{e}nyi (ER) random graphs and 
	Barab\'{a}si-Albert (BA) scale-free graphs. We also show the 
	connection between DD distribution and GA. Thereafter, in the 
	\textit{Computational results} section, we present our 
	numerical results for the DD distribution in diverse synthetic  
	and real networks. We also report our computations showing the 
	importance of DD for topological robustness in networks. 
	Lastly, we conclude with a summary and future outlook.
	
	
	\section*{Analytical results}
	
	Based on the definition of DD for an edge $e=\{v,u\}$ in an
	undirected and unweighted network given by Eq. 
	\ref{eq:degdiff_unweighted}, the probability mass function  
	$P_{\daleth}$, where $P_{\daleth}(d) = Pr(\daleth(e) = d)$, 
	is given by
	\begin{align}
		P_{\daleth}(d) &= Pr \big( | \text{deg}(v) - \text{deg}(u) | = d ~ \big| ~ \{v, u\} \in E \big) \nonumber \\
		&= \sum_{\substack{\{v, u\} \in E \\ s.t. |k-l| = d}} 
		Pr \big( \text{deg}(v)=k ~ , ~ \text{deg}(u)=l ~ | ~ \{v, u\} \in E \big).
		\label{eq:pdf_Fric_probsSum}
	\end{align}
	We next derive the analytical formulae for the DD distribution 
	in two widely-used network models. These formulae will express 
	DD distribution as a sum for Erd\"{o}s-R\'{e}nyi (ER) random 
	graphs \cite{Erdos1961} and Barab\'{a}si-Albert (BA) scale-free 
	networks \cite{Barabasi1999}. Thereafter, we present our 
	analytical calculations unravelling the connection between DD, 
	LNA and GA in undirected networks.
	
	
	\subsection*{DD distribution for Erd\"{o}s-R\'{e}nyi model}
	
	In an ER random graph, $G(n,p)$, where 
	$n$ is the number of vertices and $p$ is the probability that 
	an edge exists between any pair of vertices, the degrees of two 
	neighbouring vertices are uncorrelated except for the edge that 
	is connecting them. Therefore, the relevant quantity is the 
	\textit{excess degree} of a vertex 
	and we denote its probability distribution by $q_{k} := 
	Pr(\text{exdeg}(v)=k) = Pr(\text{deg}(v)=k+1)$. For a pair of vertices $v$ and 
	$u$ connected via an edge $\{v,u\}$, we then have the identity 
	\cite{Newman2002}
	\begin{align}
		Pr(\text{deg}(v)=k, \text{deg}(u)=l ~ | ~ \{v, u\} \in E) ~ &= ~ q_{k-1} \ q_{l-1}  .
		\label{eq:prob_fric_randomgraph}
	\end{align}
	The above relation holds since conditional on the existence of 
	an edge between two vertices their excess degree distributions 
	are independent. For given degree and excess degree 
	distributions in $G(n, p)$ \cite{Newman2010}, Eqs. 
	\ref{eq:pdf_Fric_probsSum} and  \ref{eq:prob_fric_randomgraph} 
	imply
	\begin{align}
		P_{\daleth}(d) ~ &= ~ \sum_{|k - l| = d} q_{k-1} \ q_{l-1} \nonumber \\
		&= ~ \sum_{|k - l| = d} B_{k-1}^{n-2} \ B_{l-1}^{n-2} 
		\ p^{k+l-2} \ {(1-p)}^{2n-2-(k+l)} \nonumber \\ 
		&= ~ (2 - \delta_{d, 0}) \ p^{d-2} \ (1-p)^{2(n-1)-d} ~ 
		\sum_{l = 1}^{n-1-d} \ B_{d+l-1}^{n-2} \ B_{l-1}^{n-2} 
		\ \big( \cfrac{p}{1-p} \big)^{2l} 
		\label{eq:prob_dd_exactRandomGraph}
	\end{align}
	where $B^n_k$ denotes the binomial coefficient $\binom{n}{k}$ 
	and $\delta_{d, 0}$ is the Kronecker delta, which we use to 
	avoid double counting the same permutation of $(k, l)$ when 
	$d=0$.
	
	As $n\rightarrow{+\infty}$, the degree distribution for the 
	graph ensemble $G(n,p)$ with average degree $c~=~p(n-1)$ 
	becomes the Poisson distribution 
	\cite{Newman2010}
	\begin{align}
		p_k &= e^{-c} \frac{c^k}{k!}. 
		\label{eq:poisson_random_graph} 
	\end{align}
	where $p_k$ is the probability that a given vertex has degree 
	$k$. As $q_k = \frac{(k+1)p_{k+1}}{c}$ for ER random graphs, 
	the excess degree distribution is given by 
	\begin{align}
		q_{l-1} &= e^{-c} \frac{c~^{l-1}}{(l-1)!} .
		\label{eq:exeg_dist_poisson_randomgraph}
	\end{align}
	Inserting this in Eq. \ref{eq:prob_dd_exactRandomGraph}, for 
	sufficiently large ER random graphs, we can approximate DD 
	distribution by
	\begin{align}
		P_{\daleth}(d) ~ &= ~ \sum_{|k - l| = d} e^{-2c} \cfrac{c~^{k-1}}
		{(k-1)!} \ \cfrac{c~^{l-1}}{(l-1)!} \nonumber \\ 
		&= ~ (2 - \delta_{d, 0}) \ e^{-2c} \ c^{d-2} \ 
		\sum_{l = 1}^{n-1-d} \cfrac{c^{2l}}{(d+l-1)! \ (l-1)!} .
		\label{eq:prob_DD_poissonRandomGraphs}
	\end{align}
	
\begin{figure}[htb]
		\includegraphics[width=.81\columnwidth]{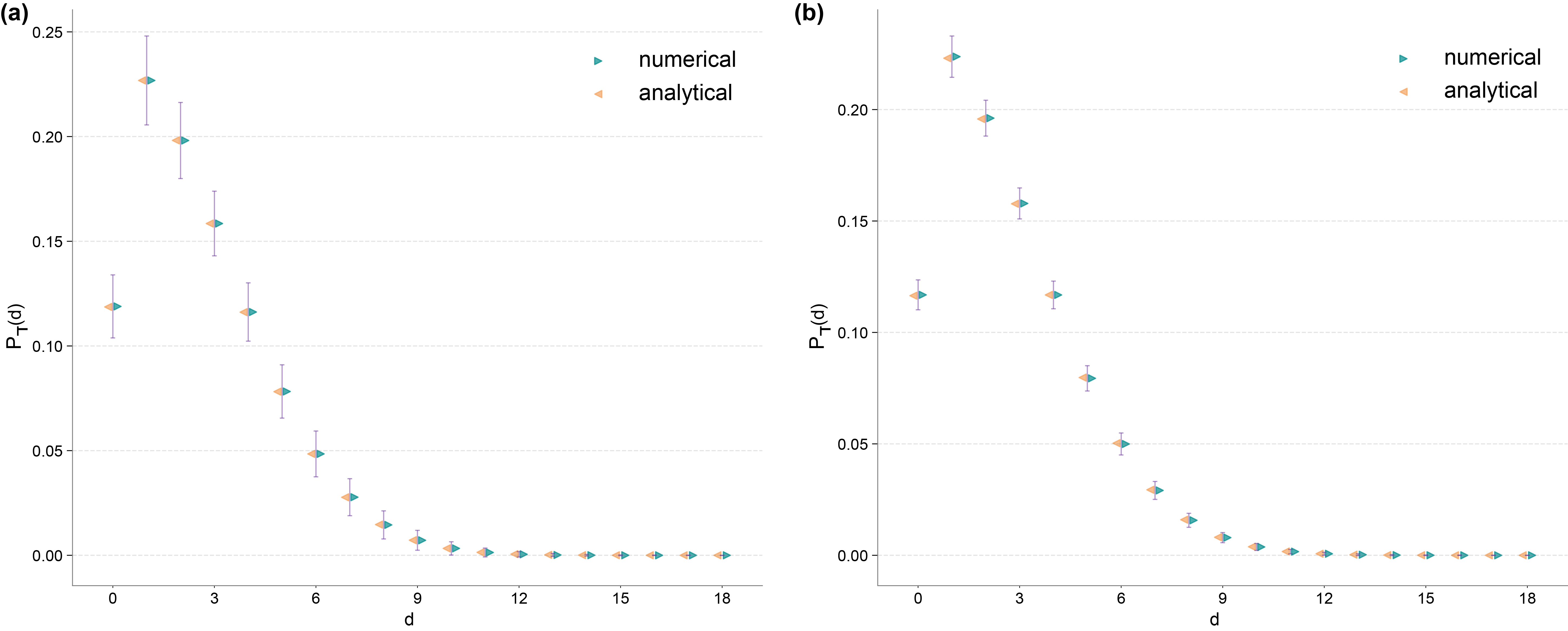}
		\caption{Concordance between analytical formula for DD 
			distribution and numerical computations in Erd\H{o}s-R\'enyi 
			(ER) model. (a) Computations for ER graph with $200$ vertices 
			and average degree $\sim 6$. In this case, Eq. 
			\ref{eq:prob_dd_exactRandomGraph} is used to obtain the 
			analytical result. (b) Computations for ER graph with $1000$ 
			vertices and average degree $\sim 6$. In this case, Eq. 
			\ref{eq:prob_DD_poissonRandomGraphs} is used to obtain the 
			analytical result. In both cases, the numerical result is 
			obtained via averaging the computed $P_{\daleth}$ over an 
			ensemble of $10000$ ER networks with same size and average 
			degree. Note that $P_{\daleth}(d)$ is of the order of $10^{-6}$ 
			or smaller for $d>18$, thus given our ensemble size, the 
			comparison between the numerical and analytical computations is 
			not valid, and thus not shown, for $d>18$.}
		\label{fig3}
	\end{figure}

	In Figure \ref{fig3}, we verify that the formulae given by Eqs.
	\ref{eq:prob_dd_exactRandomGraph} and 
	\ref{eq:prob_DD_poissonRandomGraphs} match with the numerical 
	computations for values of $d$ where $P_{\daleth}(d)$ is sufficiently 
	large considering the ensemble size.
	
	
	\subsection*{DD distribution for Barab\'{a}si-Albert model}
	
	To derive the DD distribution in a BA 
	network from Eq. \ref{eq:pdf_Fric_probsSum}, we use a
	result by Fotouhi \& Rabbat \cite{Fotouhi2013} for the joint 
	degree distribution of neighbouring vertices in a BA network 
	with $n \rightarrow{} \infty$, and this result is
	\begin{align}
		Pr\big( \text{deg}(v)=k, \  \text{deg}(u)=l \ | \  \{v, u\} \in E \big) ~ 
		&= ~ ~ \cfrac{2 \ \beta \ (\beta + 1)}{k \ (k+1) \ l \ (l+1)} 
		~ \Bigg[ 1 ~ - ~ B_{\beta + 1}^{2 \beta + 2} \ 
		\cfrac{B_{l - \beta}^{k + l - 2\beta}}{\ B_{l+1}^{k+l+2} \ } 
		~ \Bigg]
		\label{eq:jointDegreeProb_BA_0}
	\end{align}
	where $\beta$ gives the number of edges attached to the new
	vertex added at each iteration of the BA model implementing a
	preferential attachment scheme. Thereafter, using Eq. 
	\ref{eq:pdf_Fric_probsSum}, we can obtain the following 
	analytical formula for the DD distribution in BA networks 
	\begin{align}
		P_{\daleth}(d) ~ &= ~ \sum_{|k - l| = d} \ ~ \cfrac{2 \ \beta \ 
			(\beta + 1)}{k \ (k+1) \ l \ (l+1)} ~ 
		\Bigg[ 1 ~ - ~ B_{\beta + 1}^{2 \beta + 2} \ 
		\cfrac{B_{l - \beta}^{k + l - 2\beta}}{\ B_{l+1}^{k+l+2} 
			\ } ~ \Bigg].
		\label{eq:fric_sum_BA_0}
	\end{align}
	
	
	\subsection*{Connection with global assortativity}
	
	The connection between DD and GA is clear once the identity 
	in Eq. \ref{eq:mid_equn_lna_to_assort} is understood. For a graph 
	$G(V,E)$, the following identities explain the connection 
	between GA and DD distribution with LNA as an intermediate 
	step.
	\begin{align}
		\sigma_q ^2 \ r &= \sum_{j} \sum_{v \in V_j} \bigg( \Big[ (j+1) 
		\frac{j \bar{k}_v}{2M} \Big] - \Big[ (j+1) 
		\frac{j\mu_q}{2M} \Big] \bigg)  
		\label{eq:mid1_equn_degdiff_to_assort} \\
		&= \sum_{j} \sum_{v \in V_j} \bigg( \Big[ 
		\frac{j(j+1)}{2M} \big( \frac{\sum_{k\ge j}(k-j) - \sum_{k < j}(j-k)}{j+1} 
		+ j \big) \Big] - \Big[ (j+1) \frac{j\mu_q}{2M} 
		\Big] \bigg) \nonumber \\
		&= \sum_{j} \sum_{v \in V_j} \bigg( \Big[ \frac{j^2 
			(j+1)}{2M} + \frac{j\sum_{k \ge j}(k-j) - j\sum_{k < j}(j-k)}{2M} \Big] - 
		\Big[ (j+1) \frac{j\mu_q}{2M} \Big] \bigg) \nonumber \\
		&= \left[ \sum_{j} \bigg(N q_j \frac{j^2 (j+1)}{2M} \bigg)  
		+ \bigg( \sum_{j} \sum_{d} \Big[ \sum_{k=j \pm d} 
		\frac{|k-j|}{2M}  + \sum_{k=j+d} \frac{(j-1)(k-j)}{2M} - \sum_{k=j-d} \frac{(j+1)(j-k)}{2M}
		\Big] \bigg) \right]  
		\nonumber \\
		&  ~~~~~ -  \left[ \sum_j \sum_{v \in V_j} \big( (j+1) 
		\frac{j\mu_q}{2M} \big) \right] \nonumber \\
		&= \left[ \sum_d d P_{\daleth}(d) + \frac{1}{2M} 
		\sum_{j} \big(N q_j j^2 (j+1) + (j-1)  \sum_{k \ge j} (k-j)
		- (j+1)  \sum_{k < j} (j-k) \big)  
		\right] - \left[ \sum_j \sum_{v \in V_j} \big( (j+1) \frac{j\mu_q}{2M} 
		\big) \right] \nonumber \\
		&= {\left[ \langle d\rangle + \frac{N}{2M}  
			\langle j^2 (j+1)\rangle + \frac{1}{2M} \sum_{j} \sum_d 
			\Big( \sum_{k=j+d} \big[ d(j-1)  \big] - \sum_{k=j-d} \big[ d(j+1) \big] \Big) \right] - 
			\left[ \sum_j \sum_{v \in V_j} \big( (j+1) \frac{j\mu_q}{2M} \big) \right]}
		\label{eq:un_degdiff_to_assort}
	\end{align}
	where $V_j$ denotes the set of vertices with excess degree $j$, 
	$N$ is the number of vertices in the network, $d$ is the DD, 
	$P_{\daleth}$ is the probability mass function of DD distribution, and 
	the remaining notation is as in Eq. 
	\ref{eq:unbiased_local_assortativity}. 
	Note, the index $k$ under the summations in above equation refers to 
	the excess degree of only the neighbours of the vertex with excess 
	degree $j$, and we used this incomplete notation for brevity.
	The precise expanded notation, for $k=j+d$ for instance, is 
	$\{ ~ k \in \Xi_v ~~ | ~~ k=j+d ~ \}$, where $v$ is a vertex with 
	excess degree $j$ and $\Xi_v$ is the set of excess degrees of 
	neighbours of $v$.
	Also, note that the degree difference is the same as the excess degree 
	difference of two neighbouring vertices. 
	In Eq. \ref{eq:un_degdiff_to_assort}, the terms in the first outer 
	brackets show the connection between DD and the first term in the 
	summation in Eq. \ref{eq:mid_equn_lna_to_assort}, and the term in the 
	second outer brackets is the same as the second term in the summation 
	in Eq. \ref{eq:mid_equn_lna_to_assort}, i.e., the contribution of individual vertices to $\mu_q ^2$. The first outer brackets in Eq. \ref{eq:un_degdiff_to_assort} contains the first moment of DD 
	distribution (first term in the brackets), the third and second moments of excess degree distribution
	(the second term in the brackets), and a sum involving DD and excess 
	degree of an incident vertex (the third term in the brackets).
	In addition to explaining the connection between DD and GA, 
	Eq. \ref{eq:un_degdiff_to_assort} further clarifies that we 
	can compute GA and LNA using DD and excess degrees,
	while DD cannot be deduced from GA and LNA. We demonstrate 
	this remark in supplementary information (SI) Figure S1 where 
	we show DD distribution in an ensemble of BA networks as the 
	network is rewired to increase its GA, and in SI Figure S2 
	where we set a constraint on GA of ensembles of ER and BA 
	networks and show DD distribution after two random independent 
	rewirings.
	
	
	\section*{Computational results}
	
	We computed the DD distribution for 4 synthetic  networks 
	\cite{Erdos1961,Watts1998,Barabasi1999,Dall2002}, 
	namely, Erd\"{o}s-R\'{e}nyi (ER) model, Watts-Strogatz (WS) model,
	Barab\'{a}si-Albert (BA) model, and Random Geometric (RG)  model,
	and 10 empirical or real networks \cite{Barabasi2016,Leskovec2007,CAIDA,Song2010,Watts1998,Yu2008,Redner2004,Ebel2002,Schellenberger2010,Albert1999}. Of the 10 real networks 
	analyzed here, 6 are undirected and 4 are directed networks. 
	The full description of the network dataset is included in 
	the SI Appendix. We also use the 4 synthetic  networks 
	to analyze the relationship between DD and topological robustness 
	and to investigate the possible correlation between DD and other 
	edge-based measures.
	
	
	\subsection*{DD distribution in undirected networks}
	
\begin{figure}[htb]
		\includegraphics[width=.81\columnwidth]{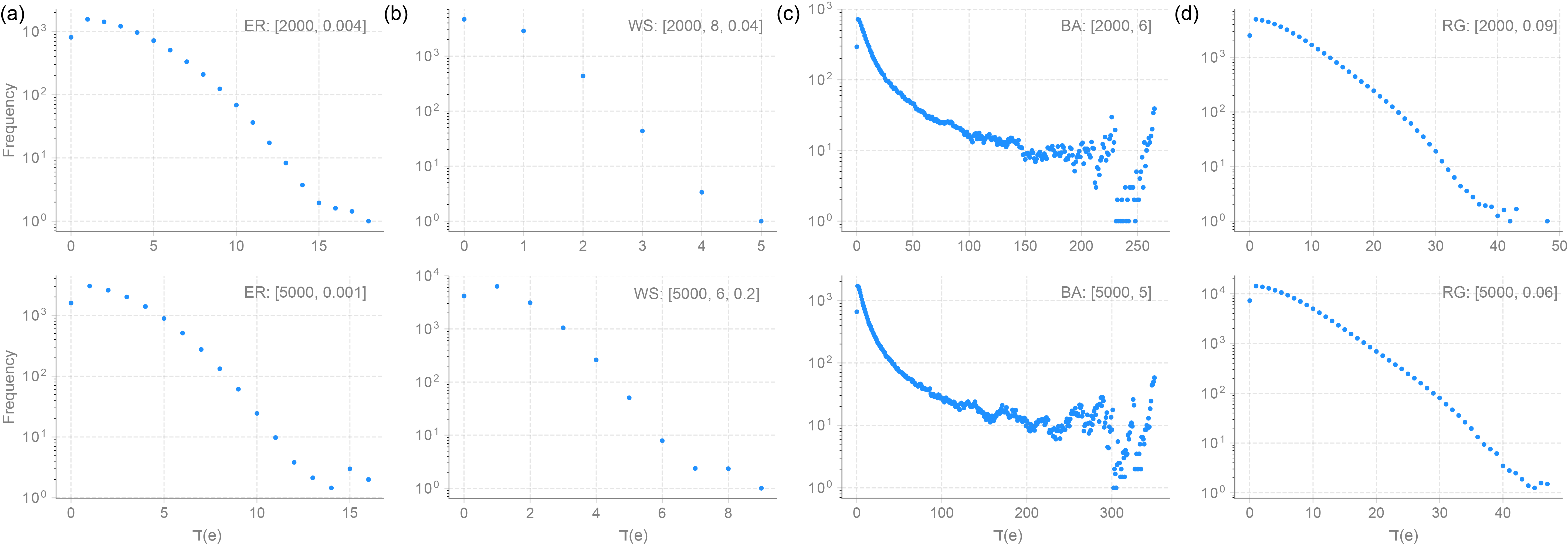}
		\caption{DD distributions in 4 different synthetic  networks. (a) 
			Erd\"{o}s-R\'{e}nyi (ER). (b) Watts-Strogratz (WS). (c) 
			Barab\'{a}si-Albert (BA). (d) Random Geometric (RG). For each 
			model, the parameters used are indicated besides it in 
			parenthesis. For ER model, the parameters are number of 
			vertices $n$ and probability $p$ of connecting an edge between 
			any pair of vertices. For WS model, the parameters are
			number of vertices $n$, the number of neighbours $k$ to 
			which each vertex is connected in the starting regular graph, 
			and rewiring probability $\beta$. For BA model, the parameters 
			are number of vertices $n$ and number of edges $\beta$ that 
			are attached to the new vertex at each iteration step. For RG 
			model, the parameters are number of vertices $n$ and radius 
			$\epsilon$. The reported correlation for each model and a given 
			set of parameters is an average over a sample of 50 networks.}
		\label{fig4}
	\end{figure}

\begin{figure}[htb]
		\includegraphics[width=.81\columnwidth]{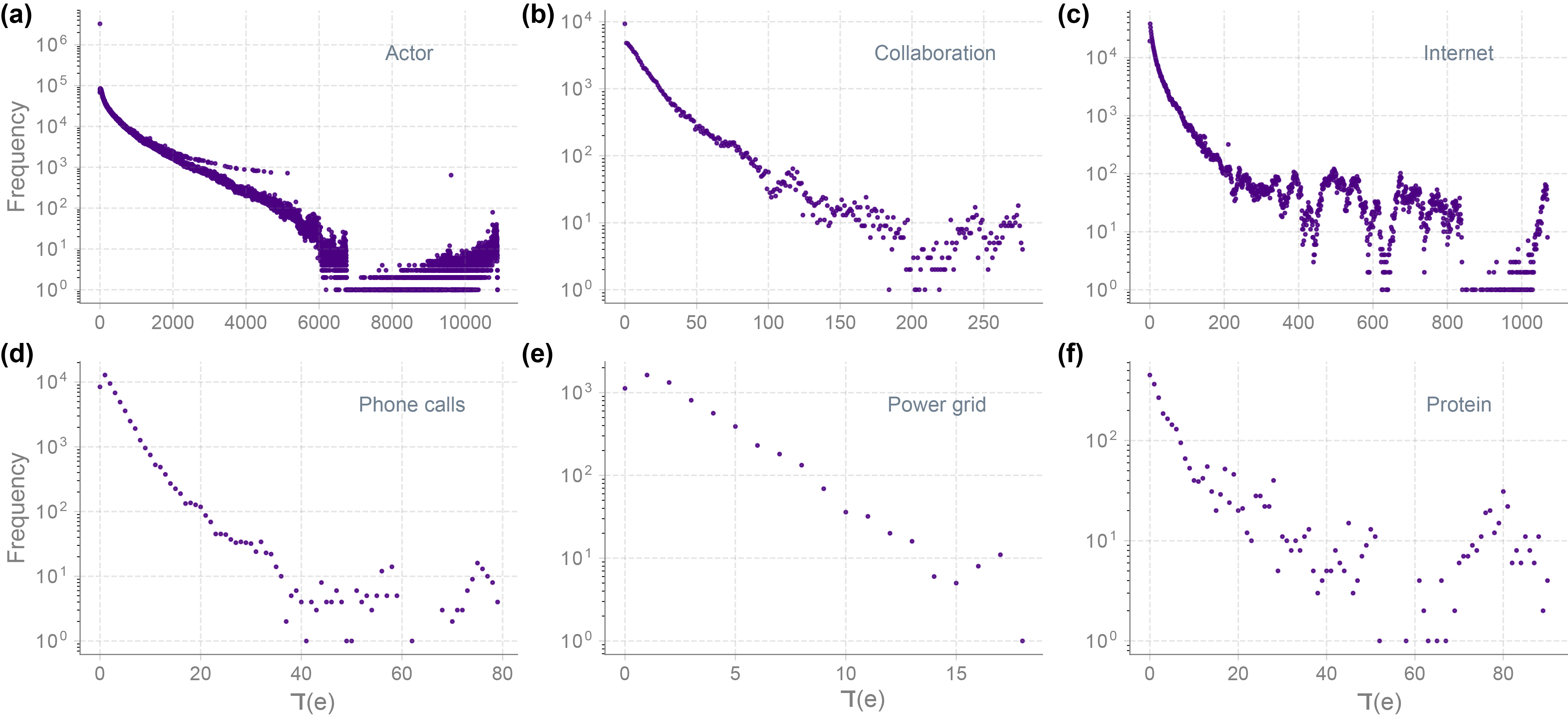}
		\caption{DD distributions in 6 undirected real-world networks. 
			(a) Actor. (b) Collaboration. (c) Internet. (d) Phone calls. 
			(e) Power grid. (f) Protein. }
		\label{fig5}
	\end{figure}

	We have computed the DD distribution of edges in 4 undirected 
	synthetic  networks and 6 undirected real networks listed in 
	SI Appendix (Figures \ref{fig4} and \ref{fig5}). From these 
	figures, we can observe qualitative differences between the DD
	distribution in different undirected networks. As the DD 
	distributions in Figure \ref{fig4} suggest, different types of 
	synthetic  networks have distinct DD distributions. In particular, 
	random geometric (RG) graphs are known to show degree assortativity 
	\cite{Antonioni2012} and the RG graphs in our dataset are 
	highly assortative with assortativity $\sim 0.55$. However, ER 
	graphs have degree assortativity close to $0$. The similarity 
	between the DD distributions in RG and ER graphs reveals a 
	remarkable fact about these two synthetic  networks; while they 
	differ significantly in GA, the mixing patterns are strikingly 
	similar at the local scale.
	
	\begin{figure}[htb]
		\includegraphics[width=.81\columnwidth]{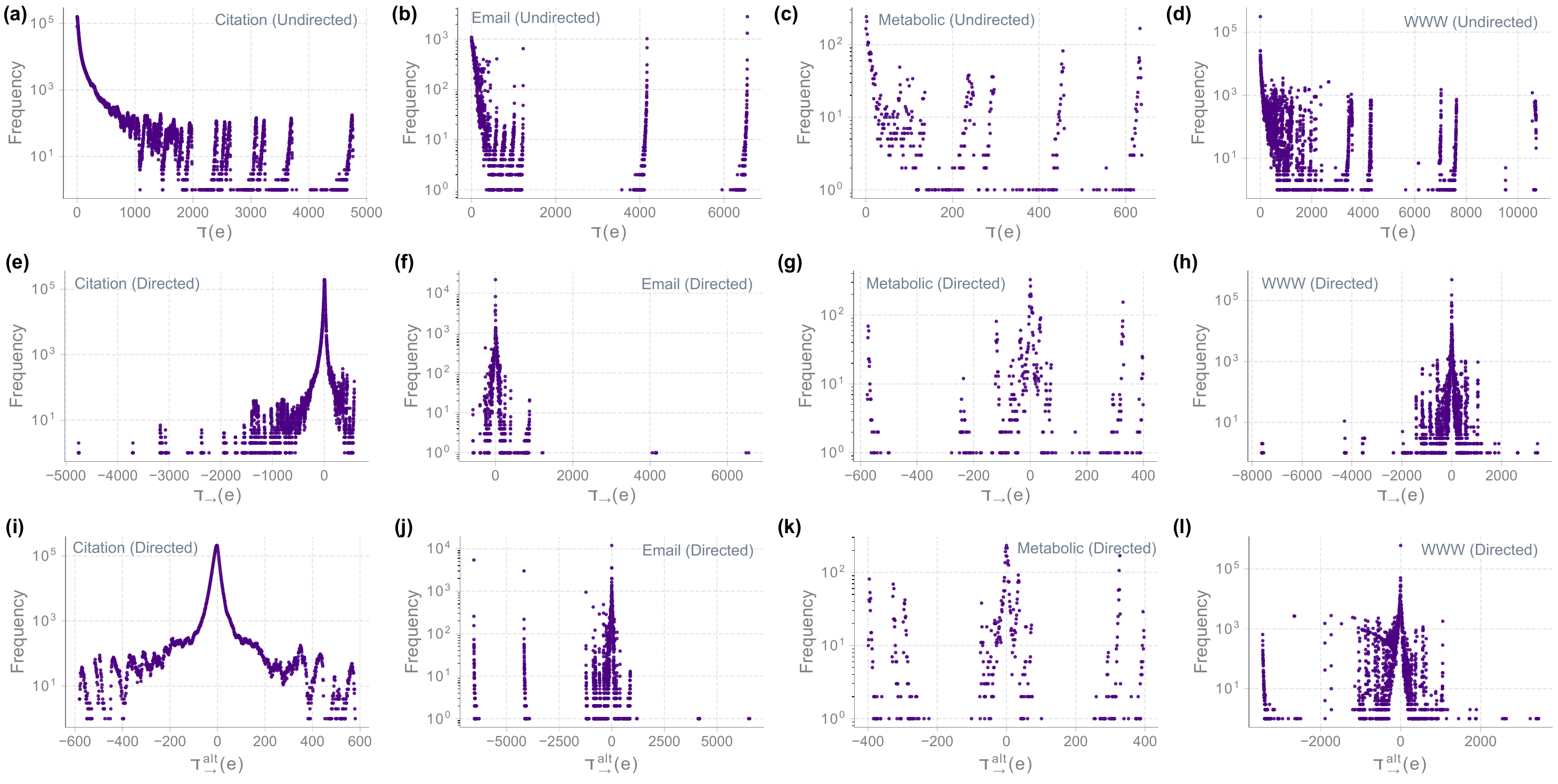}
		\caption{DD distributions in 4 directed real-world networks.
			(a-d) Distributions $\daleth(e)$ computed using Eq. 
			\ref{eq:degdiff_unweighted} by ignoring the directions of the 
			edges . (e-h) Distributions $\daleth_{\rightarrow{}}(e)$ computed 
			using the directed definition Eq. \ref{eq:di_degdiff_unweighted}.
			(i-l) Distributions $\daleth^{alt}_{\rightarrow{}}(e)$ computed 
			using an alternative definition of diDD.}
		\label{fig6}
	\end{figure}
	\subsection*{DD distribution in directed networks}
	
	DD distribution can be computed in directed networks by 
	considering the networks as undirected by ignoring the
	directions on edges. Such a computation of DD distribution in 
	undirected simplifications of directed networks could still be 
	informative of existing heterogeneity as demonstrated for the
	4 directed real networks, namely, Citation, Email, Metabolic, 
	and WWW, in Figure \ref{fig6}. To better understand the details 
	of this heterogeneity though, we can use a directed variation 
	of DD that can highlight the specifics leading to such 
	heterogeneity.
	
	The directed DD (diDD) as defined in Eq. \ref{eq:di_degdiff_unweighted},
	captures the local homophily between in-degree of the tail vertex 
	and out-degree of the head vertex of a directed edge.
	One can also define diDD differently in order to study homophily 
	between other combinations of in-degree and out-degree of vertices 
	anchoring a directed edge. In fact, distributions of diDD defined with 
	respect to each combination of in-degree and out-degree can be informative 
	in their own right. For instance, we consider diDD, $\daleth_{\rightarrow{}}$ 
	as defined in Eq. \ref{eq:di_degdiff_unweighted} as well as an alternate 
	variation of diDD defined as $\daleth^{alt}_{\rightarrow{}} 
	:= \text{deg}^{out}(\text{u}) - \text{deg}^{out}(\text{v})$ for a directed 
	edge $(v, u)$. 
	The diDD distributions in Figure \ref{fig6} enable us to make the 
	following observations about the 4 directed real networks in our 
	dataset. Metabolic network has a rather symmetric homophily in the 
	direction of the reactions, for both variations of diDD. 
	On the other hand, for 
	Citation network, while distribution of $\daleth^{alt}_{\rightarrow{}}$ 
	is relatively symmetric for the negative and positive values, the 
	distribution of $\daleth_{\rightarrow{}}$ is rather asymmetric with a 
	long tail in the negative side. In the Email network, the majority of Emails are exchanged between Email addresses with similar Email traffic. 
	In the distributions for the Email network, there 
	are two other major peaks corresponding to Emails sent from Email addresses 
	sending many Emails (e.g. organizational Email addresses) to those that 
	send only few Emails (Figure \ref{fig6}j). There are also a small number of Emails sent from 
	Email addresses receiving only a small number of Emails to those sending 
	out a large number of Emails (Figure \ref{fig6}f). Moreover, according to the distributions in 
	WWW network, within the domain of University of Notre Dame, there are 
	many hyperlinks from webpages that have links to many other webpages 
	(e.g. a departmental webpage) to those that do not contain many hyperlinks 
	to other webpages (e.g. a webpage corresponding to an announcement) (Figure \ref{fig6}l).
	
	
	\begin{figure}[htb]
		\includegraphics[width=.81\columnwidth]{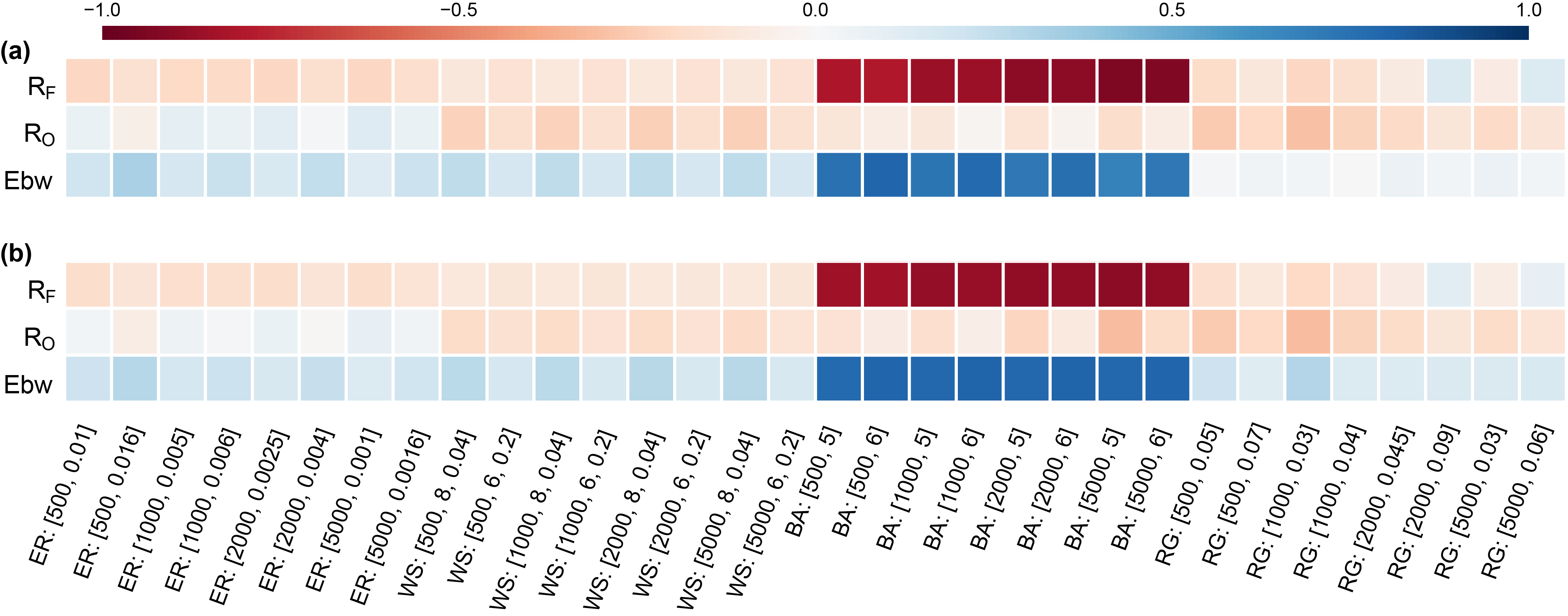}
		\caption{Correlation between degree difference (DD) and three 
			established edge-based measures, namely, Forman-Ricci curvature 
			($R_F$), Ollivier-Ricci curvature ($R_O$), and edge betweenness
			centrality (Ebw) in synthetic  networks. We show in (a) the Pearson
			correlation coefficients and in (b) the Spearman correlation 
			coefficients. For each model, the parameters used are indicated
			besides it in parenthesis. For ER model, the parameters are 
			number of vertices $n$ and probability $p$ of connecting an edge 
			between any pair of vertices. For WS model, the parameters are
			number of vertices $n$, the number of neighbours $k$ to which 
			each vertex is connected in the starting regular graph, and 
			rewiring probability $\beta$. For BA model, the parameters are
			number of vertices $n$ and number of edges $\beta$ that are 
			attached to the new vertex at each iteration step. For RG 
			model, the parameters are number of vertices $n$ and radius 
			$\epsilon$. The reported correlation for each model and a given 
			set of parameters is an average over a sample of 50 networks.
		}
		\label{fig7}
	\end{figure}
	
	\subsection*{Correlation with other edge-based measures}
	
	We explore the correlation between DD and three other 
	established edge-based measures, namely edge betweenness 
	centrality \cite{Freeman1977,Girvan2002}, Forman-Ricci 
	curvature ($R_F$) \cite{Sreejith2016,Samal2018} and 
	Ollivier-Ricci curvature ($R_O$) \cite{Jost2014,Sandhu2015,Samal2018,Ni2019} for characterizing the local network geometry. 
	These results are summarized in Figure \ref{fig7}. It is seen 
	that DD is moderately correlated with edge betweenness and 
	$R_F$ in BA networks, and this correlation is positive with 
	edge betweenness and negative with $R_F$. To avoid 
	misinterpretation, however, it is important to note that the 
	degree sum enters negatively into the definition of $R_F$ in 
	the case of unweighted and undirected graphs, due to the fact 
	that this notion originated in Riemannian geometry and 
	therefore carries over the normalizations natural in that 
	field. In SI Figure S3, we show the distribution of $R_F$ in 
	the 4 classes of synthetic  networks analyzed here. By 
	comparing with Figure \ref{fig4}, it is seen that DD provides 
	insight into the structural heterogeneity of a network, which 
	is not captured by Forman-Ricci curvature. In essence, we find that 
	degree sum and degree difference are positively correlated in 
	scale-free BA networks, which seems of interest for further 
	understanding of those networks. In general, however, one does 
	not expect such a correlation, and indeed, the correlation of 
	DD with $R_O$ in all 4 classes of synthetic  networks analyzed here, 
	and with $R_F$ in all classes other than BA networks, seems to 
	be negligible. Although edge betweenness seems to have a weakly 
	positive correlation with DD across four classes of synthetic  
	networks, this correlation seems to be noticeable only for BA 
	networks while being sufficiently small in RG graphs. 
	
	These observations further clarify that DD distribution, 
	despite its connection with measures such as edge betweenness 
	and discrete Ricci curvatures, is an independent measure. As 
	explained in other subsections, DD distribution as a 
	stand-alone measure can be informative for the local geometry 
	of the edges in the network and heterogeneity in mixing 
	patterns, and other edge-based measures considered here cannot 
	be used as a canonical proxy for DD.
	
	
	\subsection*{DD distribution and topological robustness}
	
	To test any potential relationship between DD value of edges 
	and topological robustness of the network, we here compute the 
	expected size of the largest connected component (LCC) in two 
	ensembles of ER and BA networks during reverse edge percolation 
	in increasing and decreasing order of DD. Through a comparative 
	analysis, we also investigate the importance of DD for finding 
	the minimum edge cut of the LCC.
	
\begin{figure}[htb]
		\includegraphics[width=.81\columnwidth]{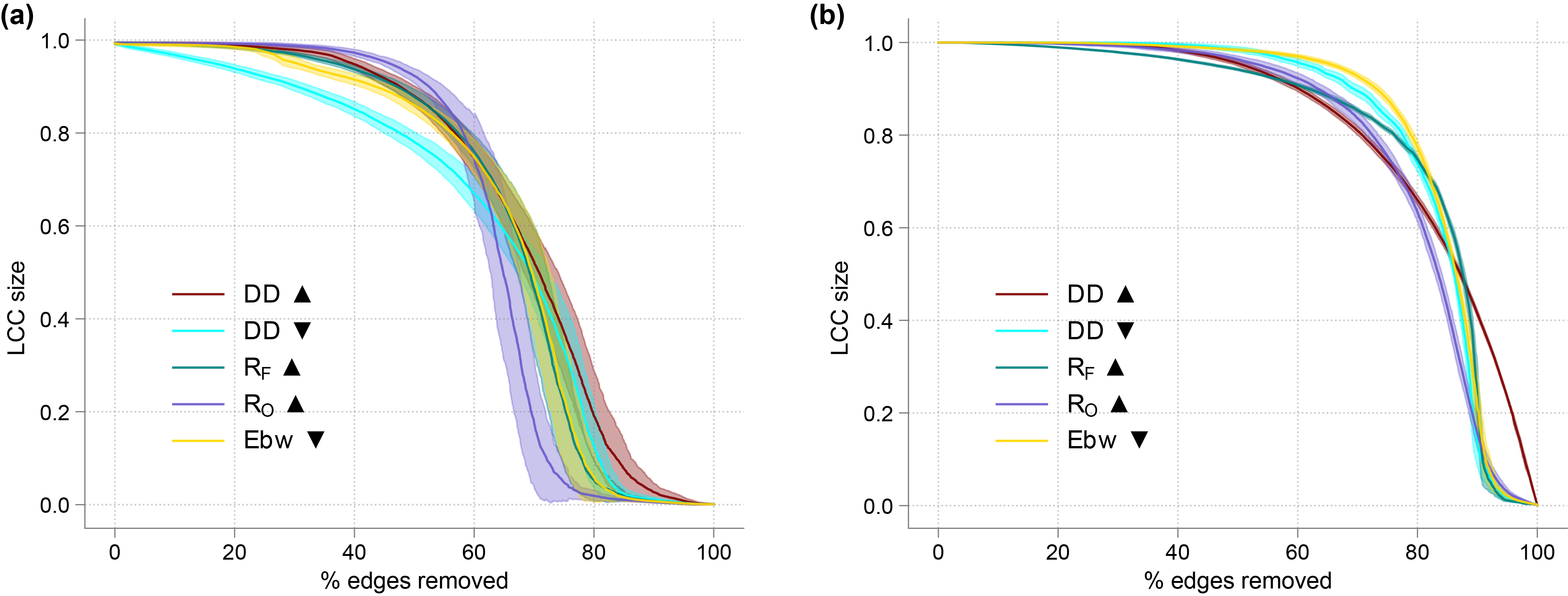}
		\caption{Topological robustness of synthetic  networks with respect to 
			deletion of edges based on DD, Forman-Ricci curvature ($R_F$), 
			Ollivier-Ricci curvature ($R_O$), and edge betweenness centrality 
			(Ebw). (a) Erd\"{o}s-R\'{e}nyi (ER) network with $n=1000$ and 
			$p=0.01$. (b) Barab\'{a}si-Albert (BA) network with $n=1000$ and 
			$\beta=5$. In each case, we show the size of the largest connected 
			component (LCC) normalized by the number of vertices in the graph 
			as a function of the percentage of edges removed. Edges are deleted 
			based on increasing and decreasing order of DD, increasing order of 
			$R_F$, increasing order of $R_O$ and decreasing order of Ebw. For 
			each model, the plots show the mean and standard deviation of LCC 
			size over an ensemble of 50 networks generated with the specified 
			parameters.}
		\label{fig8}
	\end{figure}

	Figure \ref{fig8} shows the result of this reverse edge 
	percolation analysis in ER and BA networks with respect to 
	increasing and decreasing order of DD, increasing order of 
	Forman-Ricci curvature ($R_F$), increasing order of 
	Ollivier-Ricci curvature ($R_O$), and decreasing order of edge 
	betweenness. In case of the BA network, this specific 
	simulation shows a second-order phase transition when edges 
	are removed in decreasing order of DD, a phenomenon observed 
	for edge removal in increasing order of $R_F$. Moreover, 
	in BA networks, the impact of failure of edges in decreasing 
	order of DD on LCC size seems to be only negligibly different 
	from when failure happens in decreasing order of edge 
	betweenness. This similarity in BA networks is not simply 
	due to the moderate positive correlation between these two 
	measures, but has to do with the importance of local geometry 
	for global connectivity in these networks. In other words, 
	removing edges with large DD seems to be as detrimental to 
	the LCC size as is removing edges with large edge betweenness,
	although the former, in contrast to the latter, depends only 
	on the local geometry of the network. Thus, for purposes of 
	robustness, the easily-computable and local DD can be a good 
	proxy for the global edge betweenness. Therefore, edges with 
	large DD play important roles for the global coherence in a network
	and they deserve systematic attention. 
	
	\begin{figure}[htb]
		\includegraphics[width=.81\columnwidth]{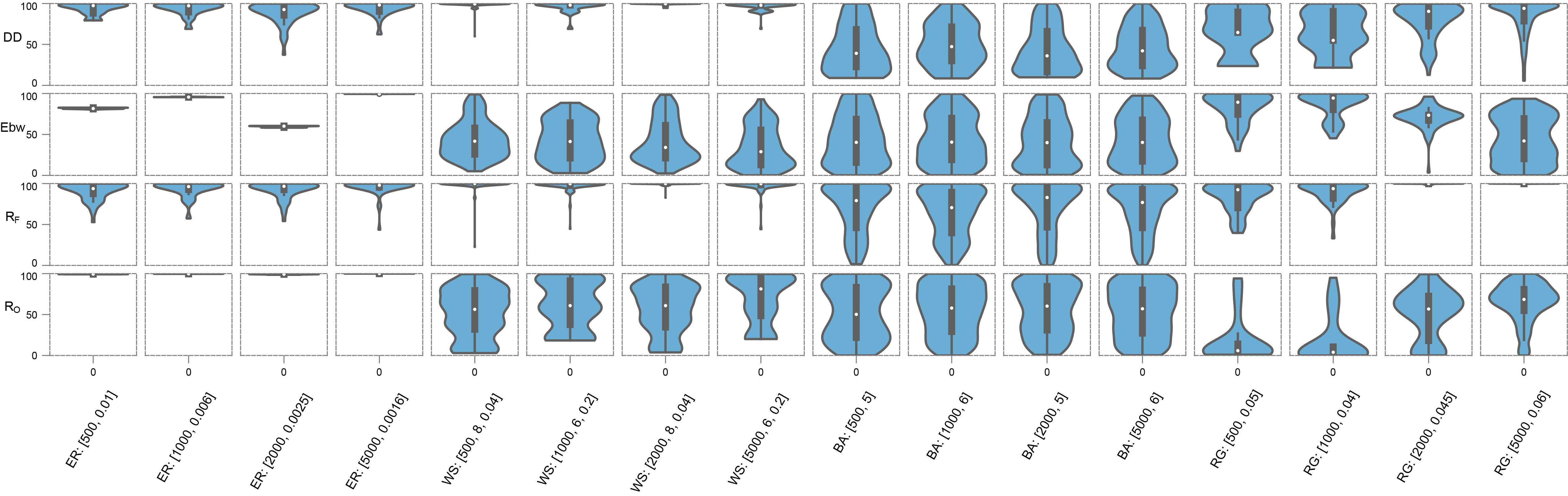}
		\caption{Comparison of the importance of the four edge-based 
			measures, DD, edge betweenness (Ebw), Forman-Ricci curvature 
			($R_F$), and Ollivier-Ricci curvature ($R_O$), towards predicting 
			the minimum edge cut (MEC) in synthetic  networks. For the MEC of the 
			LCC in each synthetic  network, we determine the percentile of each 
			edge in the MEC with respect to the 
			value of the four measures on the edges. Then, we pool
			the edge percentiles for MEC corresponding to each network in 
			the ensemble. Each violin plot shows the distribution of these 
			edge percentiles in the pool corresponding to 50 networks in the 
			ensemble for the corresponding synthetic  network. For each model, 
			the parameters used are indicated besides it in parenthesis as
			described in caption of Figure \ref{fig4}.
		}
		\label{fig9}
	\end{figure}
	
	The minimum cut \cite{Esfahanian2013} in a connected network 
	is another factor that is an indicator of topological 
	robustness. Minimum edge cut (MEC) is a set of edges of 
	minimum size that, if removed, the initially connected network 
	is no longer one connected component. In Figure \ref{fig9}, we 
	compare the importance of each of the four edge-based measures 
	-- DD, edge betweenness, $R_O$ and $R_F$ -- towards predicting 
	the MEC in the network. We compute the MEC of the LCC in each 
	of the 4 synthetic  networks analyzed here, and then, 
	determine the percentile of each edge in the MEC with respect to 
	the value of the measure on the edges in the LCC. Thereafter, we 
	pool the percentiles of the edges for MEC corresponding to each 
	network in the ensemble. In Figure \ref{fig9}, each violin plot 
	shows the distribution of these edge percentiles in the pool 
	corresponding to 50 networks in the ensemble for each synthetic  
	network. This figure shows that MEC in BA networks seems to be rather 
	uncorrelated with these edge-based measures. In RG graphs, while 
	only edge betweenness and the two curvature measures show some 
	potential for being used to infer MEC in sparser RG graphs, for 
	denser RG graphs, where most edges in the MEC seem to be almost 
	flat with respect to Forman-Ricci curvature, DD appears as the 
	second best predictor of MEC. These two best predictors of MEC 
	in dense RG graphs are, however, the worst predictors of MEC in 
	ER networks. Notably, DD seems to be the most important measure 
	for inferring MEC in Watts-Strogatz (WS) small-world networks, 
	especially when the network is highly regular. Thus, this simple 
	measure seems to play an important role in keeping the LCC connected, 
	or in other words, having a larger LCC, in a variety of network structures.
	
	
	\section*{Conclusions}
	
	Unravelling the structure of complex networks is a key interest 
	since the rise of network science. To better understand the 
	structure of large networks, it is necessary to study both the 
	global macro-scale properties and the local features from which 
	the global network structure emerges. Heterogeneity and 
	homogeneity in mixing patterns of vertices in complex networks 
	is an important known characterizing feature of network 
	structure, which reveals features beyond the degree sequence. 
	Degree assortativity was famously introduced to quantify such 
	heterogeneity at a global scale. In this contribution, we 
	study degree difference (DD) as the basic unit of mixing pattern in complex networks
	and explain the significance of this local edge-based measure. 
	We explain how this simple, elegant, computationally inexpensive
	measure can reveal valuable information about network structure. 
	A closely related measure, namely degree-degree distance, which is 
	defined as the difference of the logarithm of the degrees of both 
	vertices constituting an edge, has been recently introduced and 
	used to study scale-free property in complex networks \cite{Zhou2020}.
	In this work, we systematically explored DD as a measure for 
	structural analysis of complex networks and discuss the 
	mathematical connection between DD and global assortativity.
	Note that degree-degree distance \cite{Zhou2020}, unlike DD 
	studied here, is less closely related to assortativity. 
	Furthermore, we show that DD can be used to  
	characterize the local network geometry and shed light on an 
	understudied source of similarities or differences between
	different classes of synthetic  and real networks. 
	Notably, our numerical and analytical computations speak to 
	independence and usefulness of this measure in its own right, 
	as well as its importance for topological robustness of networks. 
	In conclusion, we recommend the simple measure, degree difference, 
	to be included in the standard toolkit of network science.
	
	Moving forward, we expect this research will seed additional 
	studies on local structural properties of complex networks in 
	both theoretical and applied settings. As the transition from 
	local to global mixing patterns in complex networks is yet to 
	be systematically explored, we believe further theoretical and 
	empirical studies on heterogeneity in mixing patterns at 
	various scales of coarse-graining can help improve our 
	understanding of the mesoscale network structure and how the 
	global topology of complex networks emerges from its local 
	geometry.
	
	
	\section*{Appendix}
	
	In this paper, we studied the DD distribution in a number of synthetic 
	and real-world networks. Moreover, we have used this dataset 
	for an empirical analysis of the properties of DD measure, its 
	significance in revealing topological features of networks, and 
	its correlation with other network measures. 
	
	\noindent The synthetic networks considered in our analysis are as follows:
	\begin{itemize}
		\item \textbf{Erd\"{o}s-R\'{e}nyi (ER) random graphs 
			\cite{Erdos1961} :} This model generates network with $n$ 
		vertices and between any given pair of vertices, there exists an 
		edge with probability $p$. 
		\item \textbf{Watts-Strogatz (WS) small-world graphs 
			\cite{Watts1998} :} This model generates network by starting 
		with a $k$-regular (ring) lattice with $n$ vertices, and then,
		each edge is randomly rewired with probability $\beta$. The 
		model gives networks with small-world property, i.e., with 
		small average path length and high clustering coefficient.
		\item \textbf{Barab\'{a}si-Albert (BA) scale-free graphs 
			\cite{Barabasi1999} :} This model generates network via a 
		preferential attachment scheme, wherein at each step, $\beta$ new 
		edges connect a new vertex to existing vertices $v$ with 
		probability proportional to $deg(v)$. These networks display 
		power-law degree distribution and scale-free property. 
		\item \textbf{Random geometric (RG) graphs \cite{Dall2002} :} 
		This model generates network with $n$ vertices, each taking a 
		random position in a 2-dimensional Euclidean plane. Thereafter, 
		by fixing a radius parameter $\epsilon$ for the network, each 
		vertex $v$ is connected to all other vertices that fall inside 
		the ball $B_\epsilon(v)$ centered at $v$. 
	\end{itemize}
	
	\noindent We have also analyzed the following undirected 
	real-world networks:
	\begin{itemize}
		\item \textbf{Actor \cite{Barabasi2016}:} This is a co-stardom 
		network with 702388 actors as its vertices and 29397908 edges 
		connecting those actors who appeared in at least one movie together.
		\item \textbf{Collaboration \cite{Leskovec2007}:} Condensed 
		Matter Physics collaboration network with 23133 vertices corresponding to
		authors who authored papers posted on arXiv during the period 
		from January 1993 to April 2003. This network has 93439 
		edges with each edge between two vertices (authors) signifying
		co-authorship in at least one paper.
		\item \textbf{Internet \cite{CAIDA}:} This is network of 192244 
		routers (vertices) with 609066 connections (edges).
		\item \textbf{Phone calls \cite{Song2010}:} This network 
		captures phone calls between a sample of active cell phone 
		users. In this network, there are 36595 users represented as 
		vertices with 56853 edges between them. Two vertices are 
		connected with an undirected edge if the corresponding users 
		have at least once made a phone call to each other over the 
		observed time interval.
		\item \textbf{Power grid \cite{Watts1998}:} This network 
		represents the power grid in western states of USA. Vertices
		are power plants and edges represent direct connections 
		between power plants via a cable. In this network, there are 
		4941 vertices and 6594 edges between them.
		\item \textbf{Protein \cite{Yu2008} :} This network is a human 
		protein-protein interaction network with 2018 proteins as 
		vertices and 2930 edges which represent mutual engagement of a 
		pair of proteins in an interaction.
	\end{itemize}
	
	\noindent In addition to the above-mentioned undirected real 
	networks, we have also analyzed the following directed real 
	networks: 
	\begin{itemize}
		\item \textbf{Citation \cite{Redner2004}:} This is a network 
		of citations between 449673 papers (vertices) published in 
		APS journals. A directed edge points from a vertex $v$ to a 
		vertex $u$ if $v$ cites $u$. There are 4685576 directed edges 
		in this network.
		\item \textbf{Email \cite{Ebel2002}:} This network is based on 
		Email communications at the University of Kiel, Germany over 
		112 days. There are 57194 vertices, which are the email 
		addresses, and there is a directed link from vertex $i$ to 
		vertex $j$ if $i$ has sent at least one email to $j$. Overall, 
		there are 93090 directed edges in this network.
		\item \textbf{Metabolic \cite{Schellenberger2010}:} This is a
		network of metabolic reactions in bacterium \textit{E. coli} 
		where vertices are metabolites and directed edges are reactions 
		linking reactants to products of reactions. This network 
		contains 1039 metabolites as vertices and 4741 reactions as 
		directed edges.
		\item \textbf{WWW \cite{Albert1999}:} This is a network of 
		hyperlinks within nd.edu domain. In this network, vertices are
		webpages and there is a directed edge from a webpage $v$ to a 
		webpage $u$ if $v$ includes at least one hyperlink to $u$. This 
		network contains 325729 vertices and 1117563 edges.
	\end{itemize}
	
	\subsection*{Acknowledgements}
	
	{We thank the anonymous reviewers for their constructive 
		comments which have helped improve the manuscript.} We also 
	thank Yasharth Yadav for discussions. A.S. would like to acknowledge 
	support from the Max Planck Society, Germany, through the award of 
	a Max Planck Partner Group in Mathematical Biology. A.F. would like 
	to thank Shirin Maleki for assistance with creating Figure 1. 
	{Open Access funding provided by Projekt DEAL.}
	
	\subsection*{Author contributions}
	
	A.F., A.S. and J.J. designed the study. A.F. performed the 
	simulations. A.F., A.S. and J.J. analyzed results. A.F., A.S. and 
	J.J. wrote the manuscript. All authors reviewed and approved 
	the manuscript.
	
	
	\bibliography{networkscience}

	\newpage
	\begin{center}
		\section*{\large \bf SUPPLEMENTARY INFORMATION (SI)}
	\end{center}
	\renewcommand{\theequation}{S.\arabic{equation}}
	\renewcommand{\thefigure}{S\arabic{figure}}
	\setcounter{equation}{0}
	\setcounter{figure}{0}
	
	\begin{figure}[ht]
		\includegraphics[width=.95\linewidth]{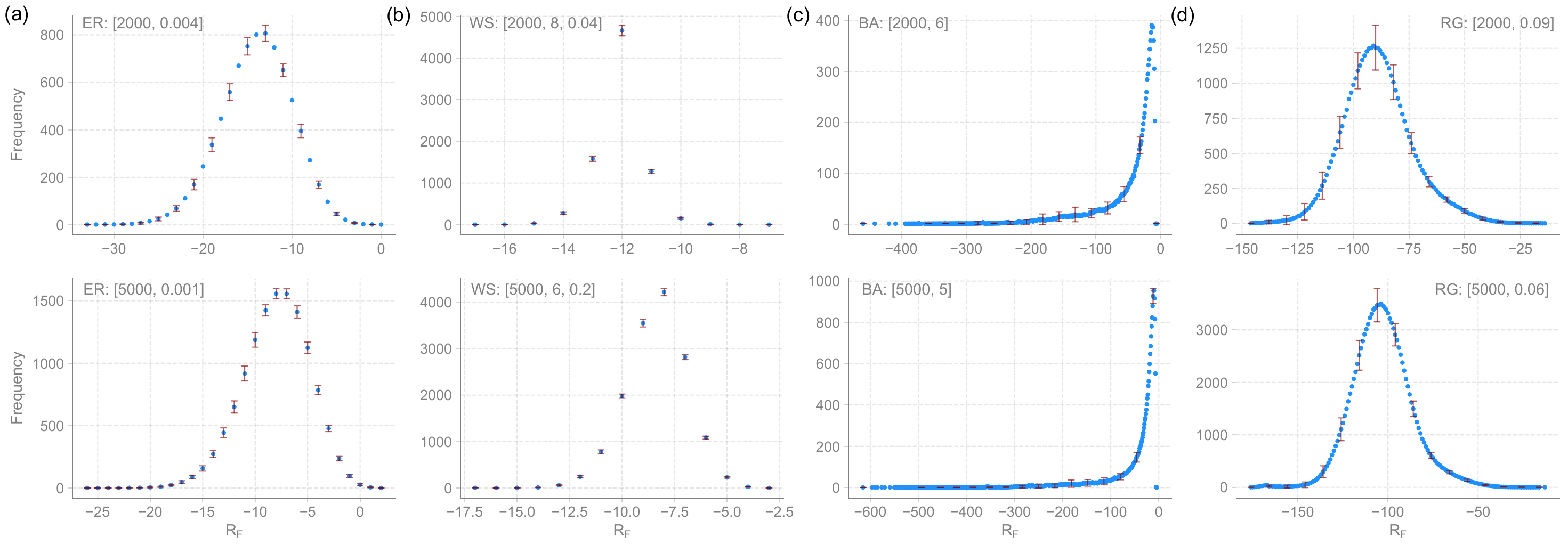}
		\caption{Forman-Ricci curvature ($R_F$) distributions in 4 
			different synthetic networks. (a) Erd\"{o}s-R\'{e}nyi (ER). (b) 
			Watts-Strogatz (WS). (c) Barab\'{a}si-Albert (BA). (d) 
			Random Geometric (RG). For each network model, the parameters used 
			are indicated besides it in parenthesis. For ER model, the 
			parameters are number of vertices $n$ and probability $p$ 
			of connecting an edge between any pair of vertices. For WS 
			model, the parameters are number of vertices $n$, the number 
			of neighbours $k$ to which each vertex is connected in the 
			starting regular graph and rewiring probability $\beta$. For 
			BA model, the parameters are number of vertices $n$ and 
			number of edges $\beta$ that are attached to the new vertex 
			at each iteration step. For RG model, the parameters are 
			number of vertices $n$ and radius $\epsilon$. The reported 
			distribution for each network model and a given set of parameters is 
			an average over a sample of 50 networks, shown as dots, 
			and the error bars show the corresponding standard deviation.}
	\end{figure}
	
	\begin{figure}[ht]
		\includegraphics[width=.95\linewidth]{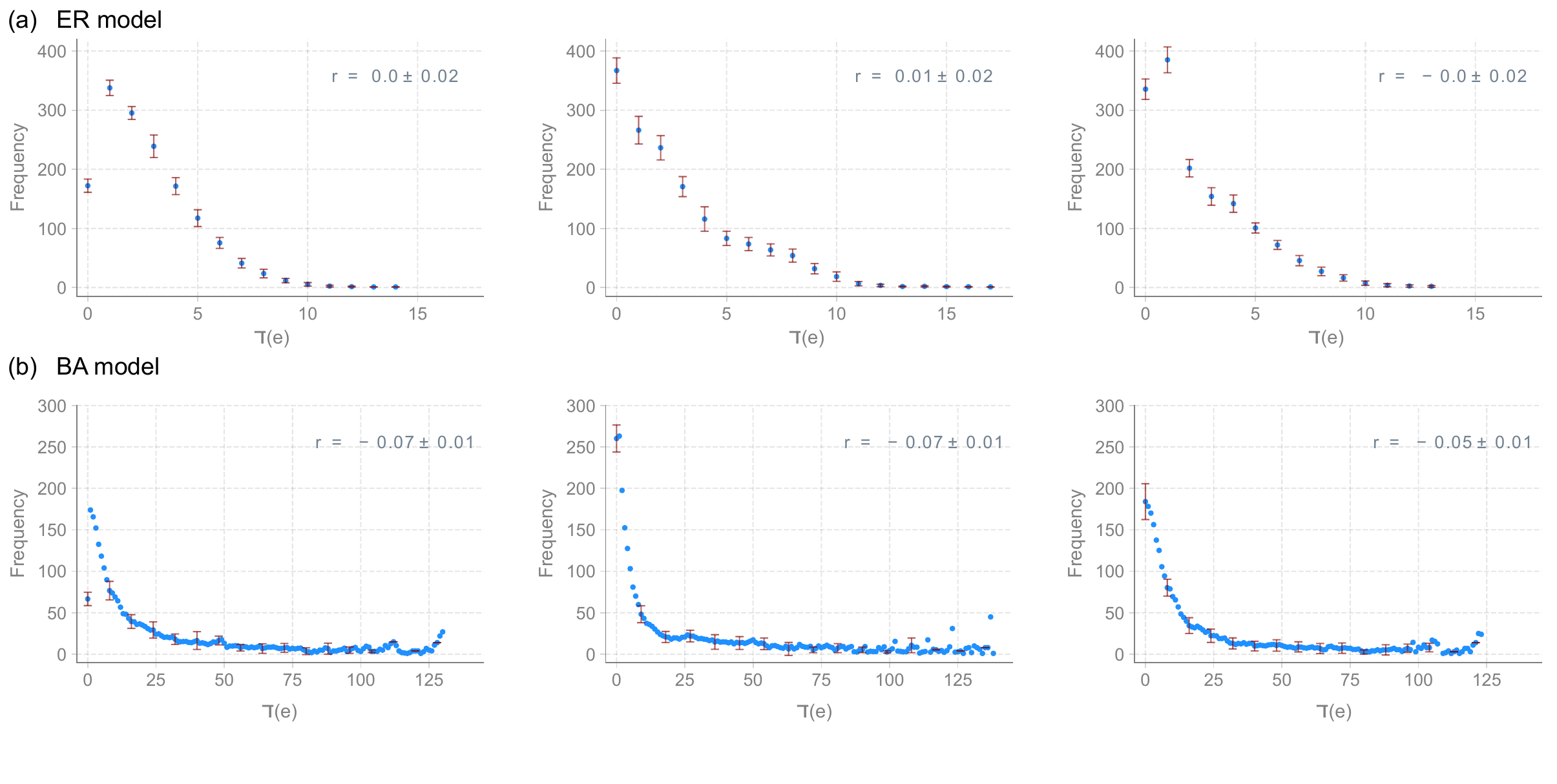}
		\caption{DD distributions for a given synthetic network and two 
			rewired networks with same degree sequence as the given network 
			and with pairwise global assortativity difference of $\le 0.025$ 
			with respect to the given (starting) network. 
			(a) Erd\"{o}s-R\'{e}nyi (ER) networks with $n=500$ and $p=0.012$. 
			(b) Barab\'{a}si-Albert (BA) networks with $n=500$ and $\beta=5$. 
			In each subfigure, we show the average and standard deviation of 
			the DD values over an ensemble of 20 networks as dots and error 
			bars, respectively. In each plot, the legend gives the average 
			and standard deviation of the global assortativities for the 
			ensemble of 20 networks. Interestingly, although the global 
			assortativity of each rewired network is $\le 0.025$ different 
			from the starting network, and the difference between global 
			assortativities of the two rewirings is $\le 0.05$, difference in 
			DD distribution is clearly visible.
		}
	\end{figure}
	
	\begin{figure}[ht]
		\includegraphics[width=.95\linewidth]{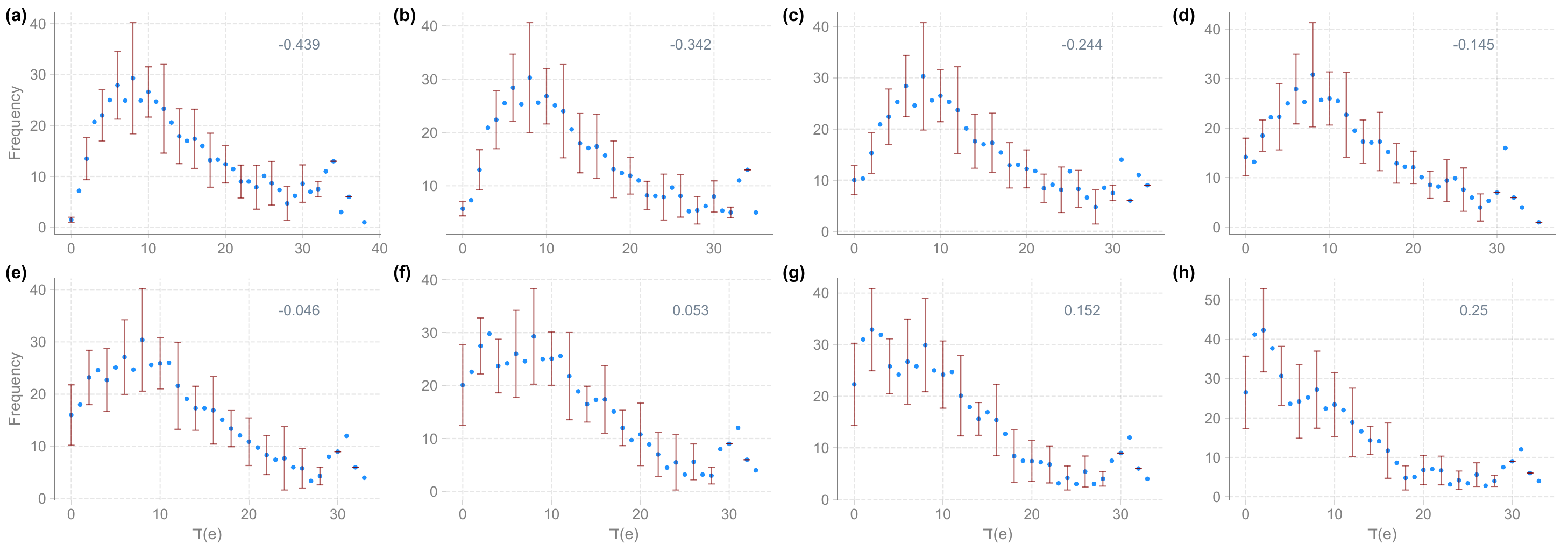}
		\caption{The evolution of DD distribution of Barab\'{a}si-Albert 
			(BA) networks created with $n=100$ and $\beta=5$ wherein 
			the degree sequence is kept fixed while global assortativity 
			differs. We start with a disassortative network and gradually increase 
			the assortativity through a targeted rewiring scheme. Briefly, this 
			heuristic to increase the assortativity of a given network is as follows.
			Given the graph at time step $t$, $G_t(V, E)$, we randomly pick two 
			edges, $\{v, u\}$ and $\{w, z\}$ from $E$. We then remove the edges 
			out of the network to obtain $\hat{G}(\hat{V}, \hat{E})$, and 
			relabel $v$, $u$, $w$ and $z$ to $v_1$, $v_2$, $v_3$, and $v_4$ 
			where the vertices are indexed in the decreasing order of their 
			degree. We next add a pair of edges $\{v_1, v_2\}$ and 
			$\{v_3, v_4\}$ to $\hat{E}$. If the assortativity of 
			$\hat{G}(\hat{V}, \hat{E})$ is greater than that of $G_t$, we 
			accept the change, and initialize $G_{t+1}$ to 
			$\hat{G}(\hat{V}, \hat{E})$, otherwise, we discard the change 
			and initialize $G_{t+1}$ to $G_t$. We continue this process for 
			a fixed number of time steps to obtain a more assortative 
			network in comparison to the starting network. In this figure, 
			we show the evolution of DD distributions for an ensemble of $20$ BA 
			networks with $n=100$ and $\beta=5$ starting as disassortative networks
			in subfigure (a) and evolving to assortative networks in (h).
			In each subfigure, we show the average and standard deviation of the 
			DD values over an ensemble of 20 networks as dots and error bars, 
			respectively. The ensemble average of the assortativity values are
			specified in the legend of each plot from (a)-(h).}
	\end{figure}
	
\end{document}